\documentclass[conference]{IEEEtran}
\IEEEoverridecommandlockouts
\usepackage{cite}
\usepackage{amsmath,amssymb,amsfonts}
\usepackage{algorithmic}
\usepackage{graphicx}
\usepackage{textcomp}
\usepackage{xcolor}
\usepackage{booktabs}
\usepackage{multirow}
\def\BibTeX{{\rm B\kern-.05em{\sc i\kern-.025em b}\kern-.08em
    T\kern-.1667em\lower.7ex\hbox{E}\kern-.125emX}}

\begin{document}

\title{Lumerical-Based SiN Half-Band FIR Maximally Flat-Top Optical Filter with Low Insertion Loss and High Extinction Ratio at 193~THz}

\author{
\IEEEauthorblockN{Md Sanowar Hossain}
\IEEEauthorblockA{\textit{School of Electrical Engineering and Computer Science} \\
\textit{University of North Dakota}\\
Grand Forks, North Dakota, USA \\
mdsanowar.hossain@und.edu}
\and
\IEEEauthorblockN{G.K.M. Hasanuzzaman}
\IEEEauthorblockA{\textit{Dept. of Electrical and Electronic Engineering} \\
\textit{Rajshahi University of Engineering and Technology}\\
Rajshahi, Bangladesh\\
gkmhasan@eee.ruet.ac.bd}
\and
\IEEEauthorblockN{Md Tanvir Mahtab Tulon}
\IEEEauthorblockA{\textit{Dept. of Electrical and Electronic Engineering} \\
\textit{Rajshahi University of Engineering and Technology}\\
Rajshahi, Bangladesh\\
tanvir.mt99@gmail.com }

\and

\IEEEauthorblockN{Md. Rakibul Islam}
\IEEEauthorblockA{\textit{Dept. of Electrical and Electronic Engineering} \\
\textit{Rajshahi University of Engineering and Technology}\\
Rajshahi, Bangladesh\\
rakibul.islam@eee.ruet.ac.bd}
}

\maketitle

\begin{abstract}
This paper presents a maximally flat-top half-band FIR optical filter on a silicon nitride (SiN) platform at 193~THz. Using a cascaded Mach--Zehnder Interferometer (MZI) topology simulated in Lumerical INTERCONNECT, MODE, and FDTD, the filter achieves insertion losses of 0.1349~dB and 0.1761~dB with extinction ratios of 18.317~dB and 23.002~dB for Channel-A and Channel-B, respectively, under realistic S-parameter conditions.
\end{abstract}
\vspace{1\baselineskip}

\section{Introduction}

Microwave photonics (MWP) combines RF engineering with photonics to enable high-frequency signal processing with low latency and wide bandwidth~\cite{capmany2007microwave, seeds2006microwave}. Optical filters are essential in MWP systems for channel selection and interference suppression, with flat-top half-band FIR filters being especially attractive for channelized signal processing~\cite{marpaung2019integrated}.

Silicon-on-insulator (SOI) has been widely used for high-order filters but suffers from two-photon absorption (TPA) and free-carrier absorption (FCA) at high optical powers~\cite{bogaerts2012silicon, dong2010low}. Silicon nitride (SiN) overcomes these issues with propagation loss below 0.1~dB/cm, broad transparency, and no nonlinear absorption, making it well-suited for C-band filtering~\cite{bauters2011ultra, moss2013new}.

Prior flat-top SiN filters using ring resonators~\cite{xu2005microring}, cascaded MZIs~\cite{khan2010reconfigurable}, and lattice structures~\cite{wang2017design} each trade off insertion loss, extinction ratio, or passband flatness. A silicon eight-tap FIR filter showed high suppression but insertion loss $>$2~dB~\cite{zhuang2015programmable}, while SiN ring designs had limited sharpness or reconfigurability~\cite{liu2019flat}. Apodized and delay-line approaches improved flatness at the cost of design complexity~\cite{chen2020apodized, wang2021delay}.

\begin{figure}[!b]
  \centering
  \includegraphics[width=\columnwidth]{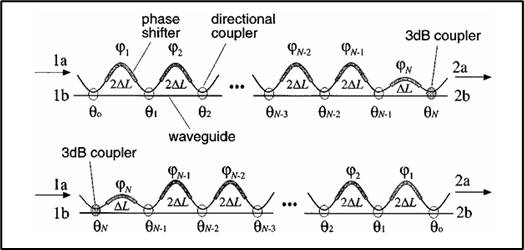}
\caption{Block diagram of the proposed cascaded MZI-based half-band FIR flat-top filter. The input optical signal propagates through $N=3$ cascaded MZI stages, corresponding to a filter order of $2N-1$. Each MZI stage is characterized by the coupling angle $\theta_k$ and phase bias $\varphi_k$, generating complementary Channel~A and Channel~B outputs.}
  \label{fig:block_diagram}
\end{figure}

This work proposes a fifth-order SiN half-band FIR maximally flat-top filter using a cascaded MZI topology with analytically tuned coupling coefficients, targeting insertion loss $<$1~dB and extinction ratio $>$20~dB at 193~THz. The overall system architecture is illustrated in Fig.~\ref{fig:block_diagram}, showing the signal flow from input through the cascaded MZI stages to the two output channels~\cite{822800}.

\section{FIR Filter Implementation}

\subsection{Specifications}

The filter is centered at 1550~nm with a free spectral range (FSR) of 100~GHz ($\approx$0.8~nm), compatible with DWDM channel spacing. A fifth-order FIR design is chosen for its unconditional stability, linear phase response, and sufficient spectral selectivity on the SiN platform~\cite{wang2018reconfigurable, bauters2011ultra}. Key parameters are listed in Table~\ref{tab:specs}.

\begin{table}[htbp]
\centering
\caption{Filter Specifications}
\label{tab:specs}
\begin{tabular}{ll}
\toprule
\textbf{Parameter} & \textbf{Value} \\
\midrule
Center Wavelength   & 1550~nm \\
Filter Type         & Half-Band FIR \\
FSR                 & 100~GHz ($\approx$0.8~nm) \\
Filter Order        & 5\textsuperscript{th} \\
Platform            & Silicon Nitride (SiN) \\
\bottomrule
\end{tabular}
\end{table}

\subsection{Waveguide}

\begin{figure}[!t]
  \centering
  \includegraphics[width=0.75\columnwidth]{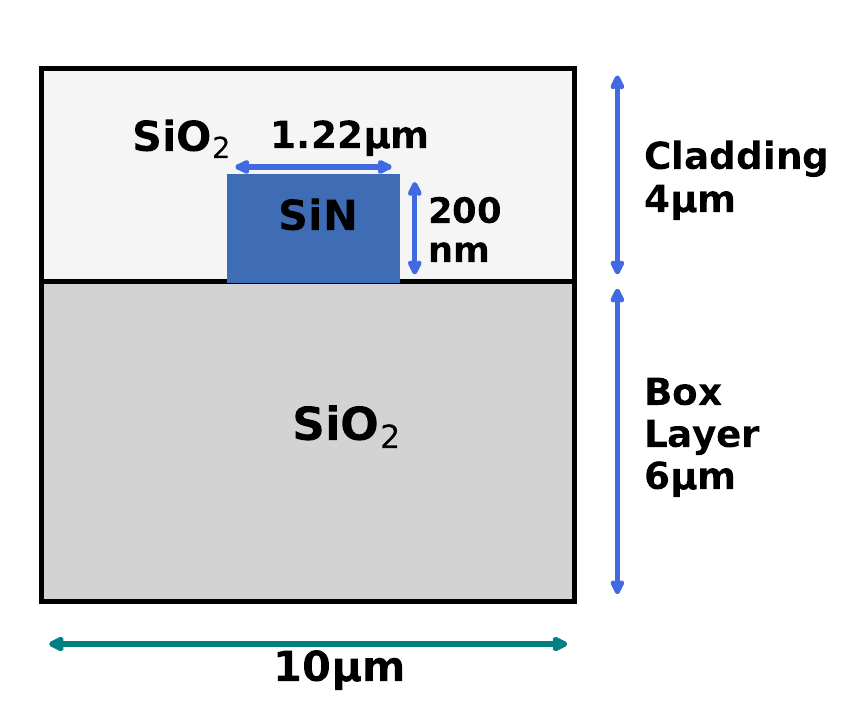}
  \caption{Cross-sectional view of the SiN waveguide: 1.22~$\mu$m~$\times$~200~nm SiN core with 4~$\mu$m SiO$_2$ upper cladding and 6~$\mu$m BOX layer~\cite{hossain2025waveguide}.}
  \label{fig:waveguide}
\end{figure}

The SiN waveguide core is 1.22~$\mu$m wide and 200~nm tall, surrounded by a 4~$\mu$m SiO$_2$ upper cladding and a 6~$\mu$m buried oxide (BOX) layer, supporting single-mode TE propagation at 1550~nm~\cite{hossain2025waveguide}. The total cross-section width is 10~$\mu$m to prevent inter-waveguide coupling~\cite{shen2020integrated}. Lumerical MODE Solutions yields $n_\text{eff} \approx 1.7801$, which is imported into INTERCONNECT for system-level consistency. The waveguide cross-section is shown in Fig.~\ref{fig:waveguide}.

\subsection{Directional Coupler}
\label{sec:dc}

\begin{figure}[htbp]
  \centering
  \includegraphics[width=\columnwidth]{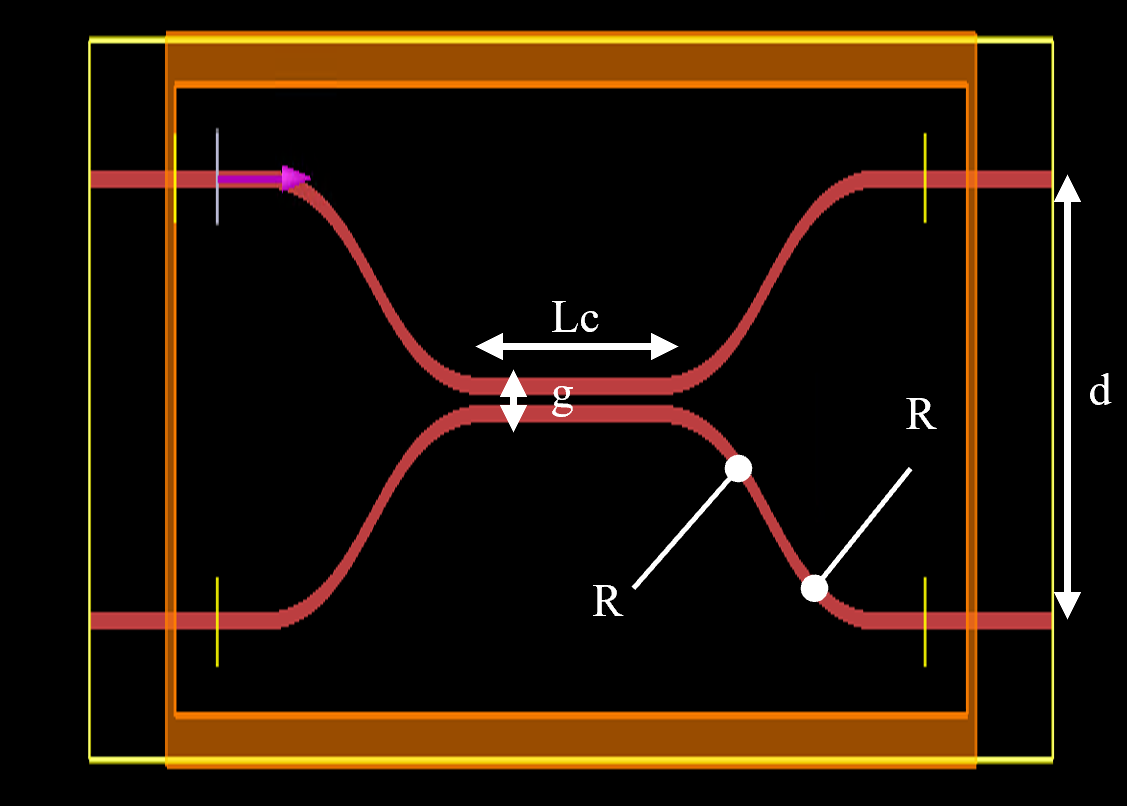}
  \caption{Schematic of the symmetric SiN directional coupler showing the two waveguides, coupling gap $g = 0.3$~$\mu$m, coupling length $L_c$, and S-bend transitions with bend radius $R = 90$~$\mu$m.}
  \label{fig:dc_block}
\end{figure}

\begin{figure}[htbp]
  \centering
  \includegraphics[width=\columnwidth,clip=true,trim=0.1in 3.4in 0.1in 3.68in]{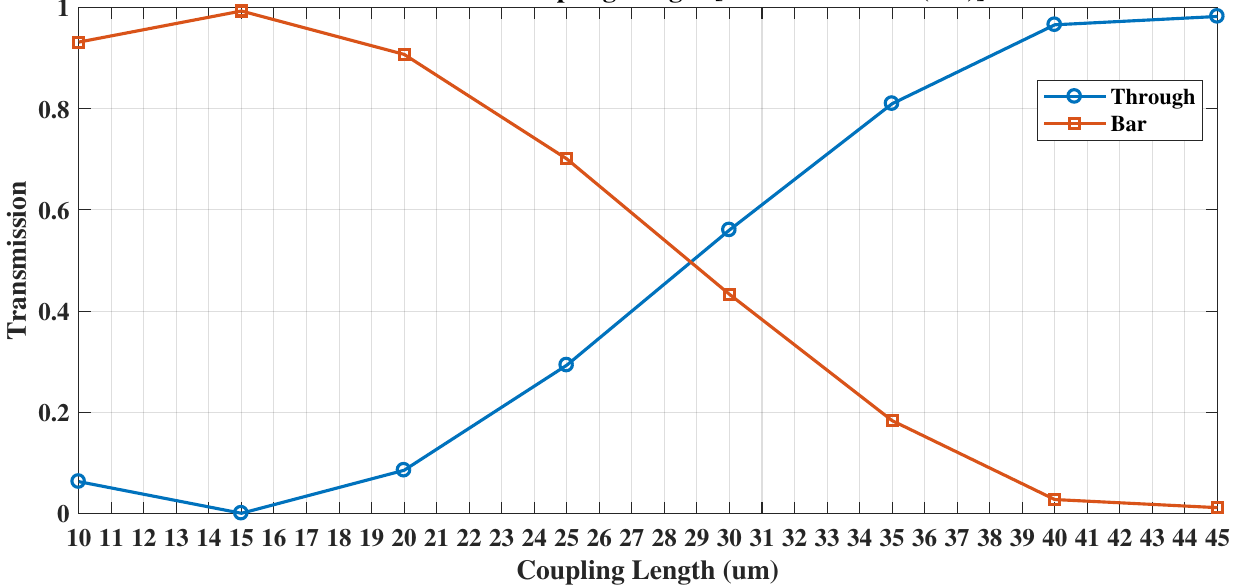}
  \caption{Simulated Through-port and Cross-port transmission as a function of coupling length $L_c$ at 1550~nm. The coupling length for each MZI stage is selected from this curve to realize the required power splitting ratio.}
  \label{fig:dc_sweep}
\end{figure}

A symmetric directional coupler with a 0.3~$\mu$m gap and 90~$\mu$m bend radius is used in each MZI stage. The coupling length $L_c$ is selected from the FDTD-simulated power transfer curve to achieve the required splitting ratio at each stage. The coupler structure is shown in Fig.~\ref{fig:dc_block}, and the simulated Through-port and Cross-port transmission as a function of $L_c$ is shown in Fig.~\ref{fig:dc_sweep}~\cite{xie2022design}. Coupler parameters are listed in Table~\ref{tab:dc}.

\begin{table}[htbp]
\centering
\caption{Directional Coupler Parameters}
\label{tab:dc}
\begin{tabular}{ll}
\toprule
\textbf{Parameter} & \textbf{Value} \\
\midrule
Coupling Length ($L_c$) & From FDTD simulation \\
Bend Radius ($R$)       & 90~$\mu$m \\
Coupling Gap ($g$)      & 0.3~$\mu$m \\
Center Wavelength       & 1550~nm \\
\bottomrule
\end{tabular}
\end{table}

\subsection{FIR Filter Design}

\begin{figure}[!t]
  \centering
  \includegraphics[width=\columnwidth]{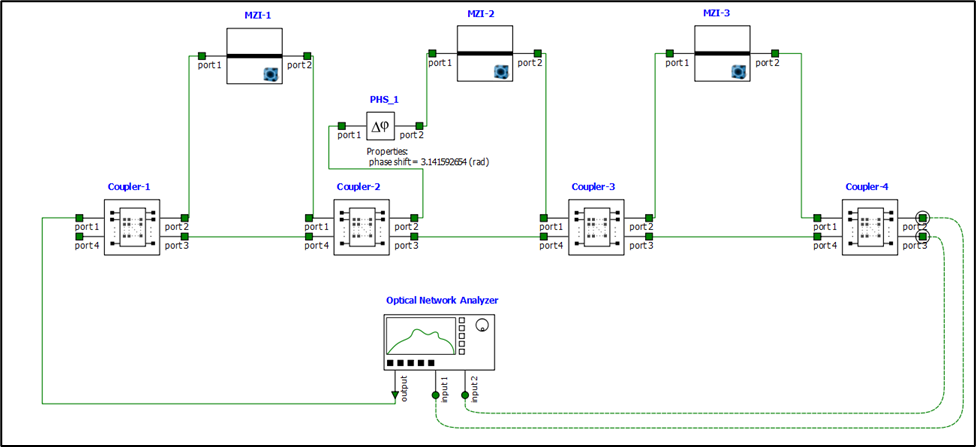}
  \caption{Lumerical INTERCONNECT schematic of the fifth-order half-band FIR filter, showing the three cascaded MZI stages with directional couplers, phase-shifting arms, and the Optical Network Analyzer for spectral characterization.}
  \label{fig:fir_interconnect}
\end{figure}

The fifth-order filter cascades three MZI stages, each with an input coupler, differential delay arms, and output coupler. The filter order follows Order $= 2N - 1$, so $N=3$ stages give a 5th-order response~\cite{wang2017design}. The analytically derived coupling angles $\theta_k$ and phase biases $\varphi_k$ are listed in Table~\ref{tab:params}. One INTERCONNECT model is built: a practical version using FDTD-extracted S-parameters to capture realistic loss and imbalance~\cite{purnawirman2013low}. The full filter schematic in Lumerical INTERCONNECT is shown in Fig.~\ref{fig:fir_interconnect}.

\begin{table}[htbp]
\centering
\caption{Circuit Parameters for Half-Band FIR Maximally Flat Filter ($N=3$)}
\label{tab:params}
\begin{tabular}{ccc}
\toprule
$k$ & \textbf{Parameter} ($x$) & $x/\pi$ \\
\midrule
0 & $\theta_0$ & 0.4664 \\
1 & $\theta_1$ & 0.1591 \\
2 & $\theta_2$ & 0.3755 \\
3 & $\varphi_0$ & 0.25 \\
4 & $\varphi_1$ & 0.50 \\
5 & $\varphi_2$ & 1.00 \\
\bottomrule
\end{tabular}
\end{table}

\section{Proposed Filter Design}

The three cascaded MZI stages implement the maximally flat condition by zeroing the first $(2N-1)$ derivatives of the transfer function magnitude at $\omega = 0$ and $\omega = \pi/2$~\cite{wang2017design, su2021optimized}. Differential arm lengths are set from $\Delta L = c/(n_g \cdot \text{FSR}) \approx 1684~\mu$m. MZI-1 and MZI-2 use $2\Delta L \approx 3368~\mu$m arms; MZI-3 uses $\Delta L \approx 1684~\mu$m. Full MZI parameters are listed in Table~\ref{tab:mzi}. In the practical model, FDTD-extracted S-parameter blocks replace ideal couplers, capturing wavelength-dependent splitting, insertion loss, and fabrication variations~\cite{su2021optimized, ding2018silicon}.

\begin{table}[htbp]
\centering
\caption{MZI Stage Parameters}
\label{tab:mzi}
\begin{tabular}{lccc}
\toprule
\textbf{Parameter} & \textbf{MZI-1} & \textbf{MZI-2} & \textbf{MZI-3} \\
\midrule
$n_\text{eff}$         & 1.7801   & 1.78033  & 1.7801  \\
$n_g$                  & 1.7801   & 1.7801   & 1.7801  \\
Arm Length ($\mu$m)    & 3368.265 & 3368.265 & 1684.133 \\
Coupling (Power)       & 0.9889   & 0.2297   & 0.8547  \\
\bottomrule
\end{tabular}
\end{table}

\section{Results and Discussion}

\begin{figure}[!t]
  \centering
  \includegraphics[width=\columnwidth,clip=true,trim=0.9in 3.42in 1in 3.73in]{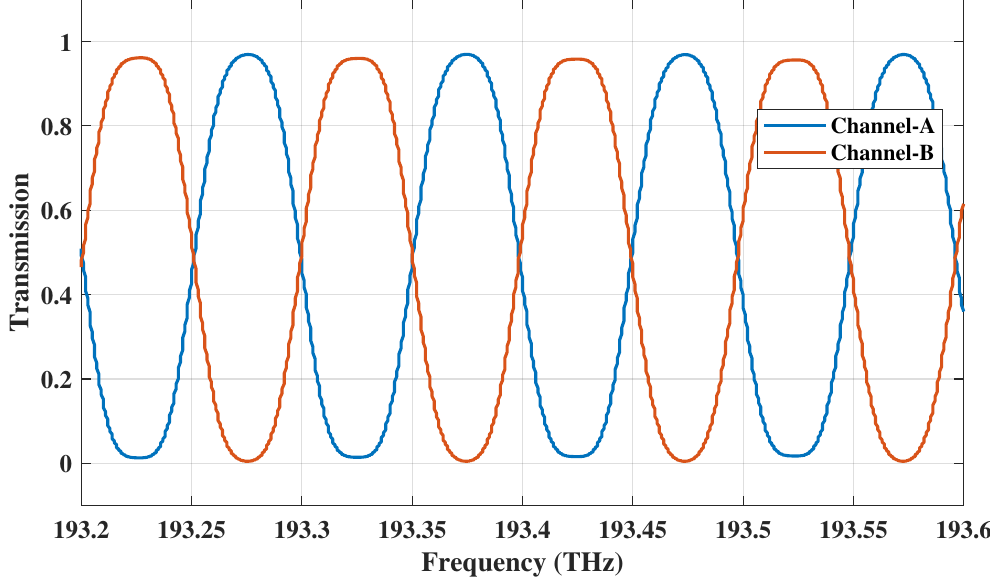}
  \caption{Transmission versus frequency for Channel-A and Channel-B across one FSR (100~GHz), illustrating the complementary half-band response and the flat-top passband characteristic of the proposed filter.}
  \label{fig:TvsF}
\end{figure}

The filter is evaluated by replacing the ideal coupler elements in the INTERCONNECT schematic with realistic S-parameter models extracted from Lumerical FDTD simulations. Fig.~\ref{fig:TvsF} shows the transmission response of both channels across the simulated frequency range. Channel-A and Channel-B clearly alternate between passband and stopband every half-FSR ($\approx$50~GHz), producing the expected complementary half-band behavior and confirming that the coupling coefficients are correctly implemented.

\begin{table}[htbp]
\centering
\caption{S-Parameter Simulation Performance Summary}
\label{tab:results}
\begin{tabular}{lcc}
\toprule
\textbf{Channel} & \textbf{Insertion Loss (dB)} & \textbf{Extinction Ratio (dB)} \\
\midrule
A & 0.1349 & 18.317 \\
B & 0.1761 & 23.002 \\
\bottomrule
\end{tabular}
\end{table}

\begin{figure}[!t]
  \centering
  \includegraphics[width=\columnwidth,clip=true,trim=0.9in 3.42in 1in 3.73in]{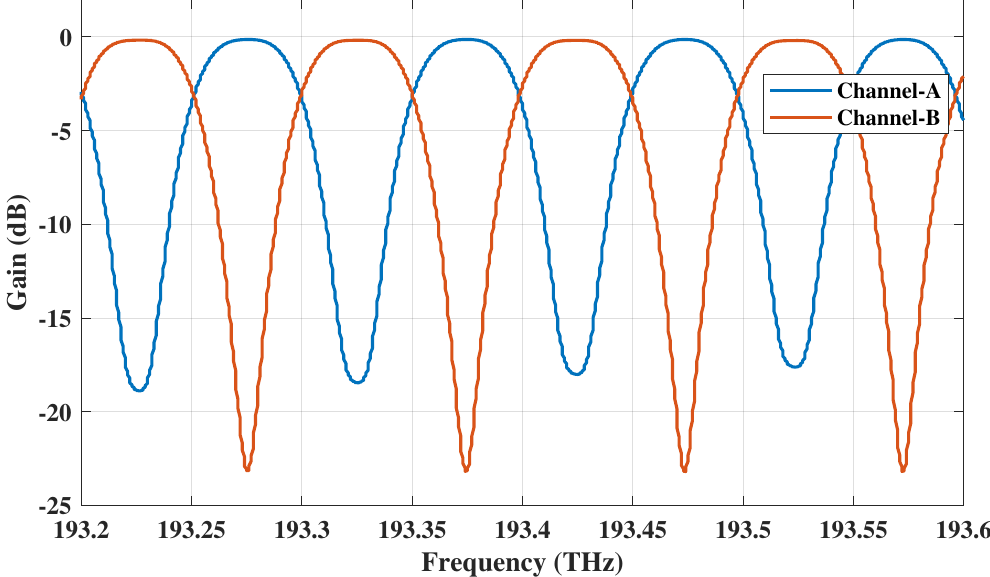}
  \caption{Practical S-parameter simulation: gain (dB) versus frequency for Channel-A and Channel-B, showing insertion losses of 0.1349~dB and 0.1761~dB with extinction ratios of 18.317~dB and 23.002~dB, respectively.}
  \label{fig:gain_db}
\end{figure}

\begin{figure}[!t]
  \centering
  \includegraphics[width=\columnwidth,clip=true,trim=0.9in 3.42in 0.65in 3.73in]{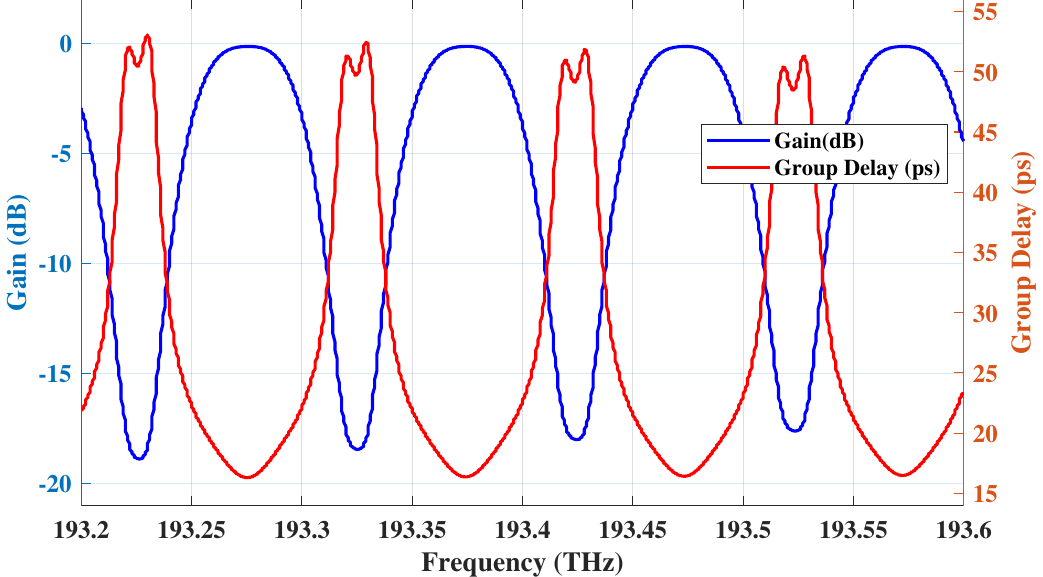}
  \caption{Gain (dB) and group delay (ps) versus frequency for the fifth-order half-band FIR filter, showing a flat passband at 193~THz with near-constant group delay across the passband.}
  \label{fig:gain_gd}
\end{figure}
The corresponding gain in dB is shown in Fig.~\ref{fig:gain_db}. Channel-A achieves an insertion loss of 0.1349~dB and an extinction ratio of 18.317~dB, while Channel-B yields 0.1761~dB insertion loss and 23.002~dB extinction ratio. Both channels retain a well-defined flat-top passband, demonstrating that the maximally flat design objective holds up under realistic S-parameter conditions. The key results are collected in Table~\ref{tab:results}.
Fig.~\ref{fig:gain_gd} overlays the gain and group delay versus frequency. The group delay stays within a range of roughly 15~ps to 50~ps and follows a smooth, predictable profile through the passband behavior that is characteristic of an FIR filter's inherently linear phase response and is important for avoiding signal distortion in MWP applications~\cite{wang2018reconfigurable}.
The extinction ratio values of 18--23~dB are more than adequate for practical DWDM filtering and RF photonic channelization~\cite{marpaung2019integrated, zhuang2015programmable}. Compared to previously reported SiN flat-top filters that either sacrifice passband sharpness~\cite{liu2019flat} or demand intricate fabrication processes~\cite{chen2020apodized, wang2021delay}, the proposed design strikes a good balance between performance and implementation simplicity.

\section{Conclusion}

A fifth-order SiN half-band FIR maximally flat-top filter at 193~THz has been designed and simulated using Lumerical MODE, FDTD, and INTERCONNECT. The cascaded MZI topology achieves insertion losses of 0.1349~dB and 0.1761~dB with extinction ratios of 18.317~dB and 23.002~dB under realistic S-parameter conditions, without requiring active tuning or complex apodization. The design is scalable and suitable for DWDM channel selection, optical channelization, and wideband RF photonic filtering. Future work will include experimental fabrication, dispersion analysis, and higher-order filter extensions.


\bibliographystyle{IEEEtran}
\bibliography{reference}

\end{document}